\documentclass[letterpaper, 12pt]{article}

\title{Growth Estimators and Confidence Intervals for the Mean of Negative Binomial Random Variables with Unknown Dispersion}
\author{David Shilane and Derek Bean}

\usepackage{graphicx}  % need this to include pictures
\usepackage{amsmath}   % i like this for the added math macros, but
\usepackage{amssymb}
\usepackage[sort&compress]{natbib}
\usepackage{geometry}
\begin{document}
\maketitle
\pagestyle{empty}

%\noindent\textbf{\large{Abstract}}

%\emph{Keywords:} Bernstein's Inequality, Confidence Intervals, Growth Estimator, Negative Binomial Distribution.

\section{Introduction}
\label{introP4}

Confidence intervals are routinely applied to limited samples of data based upon their asymptotic properties.  For instance, the Central Limit Theorem states that the sample mean $\bar X$ will have an approximately Normal distribution for large sample sizes provided that the data's second moment is finite.  This Normal approximation is a fundamental tool for inferences about the data's expected value in a wide variety of settings.  Confidence intervals for the mean are often based upon Normal quantiles even when the sample size is very moderate (e.g. 30 or 50).  However, the Normal approximation's quality cannot be ensured for highly skewed distributions \citep{Wilcox}.  In this setting, the sample mean may converge to the Normal in distribution at a much slower rate.  The Negative Binomial distribution is known to have an extremely heavy right tail, especially under high dispersion.\\

\noindent In previous work \citep{Shilane-nb-2010}, we established that the Normal confidence interval significantly under--covers the mean at moderate sample sizes.  We also suggested alternatives based upon Gamma and Chi Square approximations along with tail probability bounds such as Bernstein's Inequality.  We now propose growth estimators for the mean.  These estimators seek to account for the relative over--representation of zero values in highly dispersed Negative Binomial data.  This may be accomplished by imposing a correction factor as an adjustment to the mean or by directly removing a small number of zero values from the sample.  We will demonstrate that these alternative procedures provide a confidence interval with improved coverage.  Estimators based on the growth method can also be shown to asymptotically converge in sample size to the Normal approximation.\\

\noindent Section \ref{NBdist} reviews the Negative Binomial distribution and provides some discussion of parameter estimation under high dispersion.  Section \ref{priormethods} reviews existing methods of constructing confidence intervals for the mean of Negative Binomial random variables.  In Section \ref{growth}, we introduce the Growth Method and propose two new confidence intervals for the Negative Binomial mean.  Section \ref{sim} conducts a simulation experiment to compare these methods to existing procedures in terms of coverage probability.  Finally, we will conclude the paper with a discussion in Section \ref{conclusion}.

\section{The Negative Binomial Distribution}\label{NBdist}

\noindent The Negative Binomial distribution models the probability that a total of $k\in\mathbb{Z}^+$ failures will result before $\theta\in\mathbb{R}^+$ successes are observed.  Each count $X$ is constructed from (possibly unobserved) independent trials that each result in success with a fixed probability $p\in(0,1)$.  The expected value of $X$ is $\mu = \theta\left(\frac{1}{p}-1\right)$.  The Negative Binomial distribution may be alternatively parameterized in terms of the mean $\mu$ and \emph{dispersion} $\theta$ directly \citep{Hilbe}.  For any $\mu\in\mathbb{R}^+$ and $\theta\in\mathbb{R}^+$, a Negative Binomial random variable $X\sim NB(\mu,\theta)$ has the probability mass function

\begin{equation}\label{pmf}
P(X=x) = \frac{\mu^x}{x!} \frac{\Gamma(\theta+x)}{\Gamma(\theta)[\mu+\theta]^x}\frac{1}{\left(1+\frac{\mu}{\theta}\right)^\theta}, x\in\mathbb{Z}^+.
\end{equation}

\noindent The variance of $X$ is given by $\sigma^2=\mu + \mu^2/\theta$.  As $\theta$ grows large, Equation (\ref{pmf}) converges to the probability mass function of a Poisson random variable.  Smaller values of $\theta$ lead to larger variances, so the selection of $\theta$ controls the degree of dispersion in the data.  At very small values of $\theta$, the dispersion becomes extreme, and the data may be relatively sparse.  Because the Negative Binomial distribution has a heavy right tail, the small number of non--zero values may be spread over an extremely wide range.  In light of these concerns, it is not surprising that the Normal approximation exhibits a slow convergence as a function of sample size.\\

\noindent For all methods, we assume that the data consist of $n\in\mathbb{Z}^+$ independent, identically distributed (i.i.d.) $NB(\mu,\theta)$ random variables, where $\mu$ and $\theta$ are unknown.  We seek to generate accurate and reliable inferences about $\mu$.  The dispersion $\theta$ may be considered a nuisance parameter.  In the previous work of \citet{Shilane-nb-2010}, some methods relied upon estimates of $\theta$ while others directly estimated the variance $\sigma^2$ with the unbiased estimator $s^2$.  In general, we prefer to estimate the variance directly where possible.  A variety of research suggests that estimating small values of $\theta$ is especially difficult in small sample sizes.  Some existing procedures include the method of moments estimator $\hat\theta = \bar X / ((s^2 / \bar X) - 1)$ and an iterative maximum likelihood estimator (MLE)  \citep{Piegorsch,ClarkPerry}.  \citet{Aragonetal} and \citet{Ferreri} provide conditions for the existence and uniqueness of the MLE.  Meanwhile, \citet{Pietersetal} compares an MLE procedure to the Method of Moments at small sample sizes.  These procedures encounter difficulties when the variance estimate $s^2$ is less than the sample mean $\bar X$.  The method of moments estimator will provide an implausible negative number, while maximum likelihood procedures will produce highly variable results by constraining $s^2$ to be at least as large as $\bar X$.  (Additionally, the \textbf{glm.nb} method in the \textbf{R} Statistical Programming Language will often produce computational errors in this setting rather than return an MLE for $\theta$.)\\

\noindent For any $\alpha\in(0,1)$, we seek to construct high--quality $1-\alpha$ confidence intervals for the mean $\mu$ based upon a sample $X_1,\dots,X_n$ of i.i.d. $NB(\mu,\theta)$ random variables.  A method's \emph{coverage probability} is the chance that the interval will contain the parameter of interest $\mu$ as a function of the sample size $n$ and the parameters $\mu$ and $\theta$.  We will primarily judge the quality of a confidence interval in terms of its coverage probability, which ideally would be exactly $1-\alpha$ across all sample sizes and parameter values.  However, there are many secondary factors that can impact the selection of methods.  Shorter intervals provide greater precision and insight about the underlying scientific problem.  The variability of this length should also be minimized.  When this variability is too great, the resulting interval may significantly understate or overstate the degree of certainty about the parameter range.  Where possible, we prefer methods that can be assured of producing plausible parameter ranges.  Since the Negative Binomial distribution draws from a non--negative sample space, we therefore prefer methods that will result in a non--negative confidence interval.\\

\section{Prior Methods}\label{priormethods}

\noindent The Normal approximation and the bootstrap Bias Correct and Accelerated (BCA) method \citep{Bootstrap} are considered standard techniques for the construction of $1-\alpha$ confidence intervals for the mean.  \citet{Shilane-nb-2010} found that these procedures perform similarly over a wide variety of sample sizes and parameter values in Negative Binomial models.  They also proposed several alternative methods, including Gamma and Chi Square approximations along with tail probability bounds such as Bernstein's Inequality.  These methods improved upon the standard techniques in terms of coverage probabilities over complimentary subspaces of parameter values.  We will briefly review these techniques in the following sub--sections.

\subsection{The Gamma Approximation}\label{gammasubsectionP4}

\noindent \citet{Shilane-nb-2010} proved a limit theorem stating that $\bar X$ converges to a Gamma distribution as the sample size $n$ grows large and the dispersion parameter $\theta$ approaches zero.  The shape parameter is $\theta n$, and the rate parameter is given by $\frac{\theta n}{\mu}$.  The Gamma approximation therefore requires estimates of $\mu$ and $\theta$.  The sample mean $\bar X$ may be plugged in for $\mu$ directly.  Due to the difficulties previously discussed with MLE methods under high dispersion, we will rely upon a method of moments estimator of $\theta$ as a default.  (In practice, an MLE may be substituted where possible.)  To account for the possibility of negative values, we recommend truncating all values below a small number (such as $10^{-5}$ by default) to ensure positive estimates.  Once the shape and rate parameters are estimated, a $1-\alpha$ confidence interval for $\mu$ is given by the $(\alpha/2)$nd and $(1-\alpha/2)$th quantiles of the Gamma distribution.\\

\noindent The Gamma approximation generally performs best when $n$ is large and $\theta$ is small.  In this setting, the Gamma method improves upon the Normal approximation by a coverage probability of about 1--2$\%$.  At more moderate sample sizes and dispersions, the Gamma approximation is not especially accurate and often performs worse than the Normal approximation.

\subsection{The Chi Square Approximation}\label{chisqapproxP4}

A special case of the Gamma approximation of Section \ref{gammasubsectionP4} occurs when $\mu=2n\theta$.   In this setting, the Gamma parameters correspond to a Chi Square distribution with $\mu$ degrees of freedom.  This Chi Square approximation works especially well when the parameter relationship is approximately equivalent.  The parameter relationship may also be phrased in terms of a \emph{ratio statistic} $\mu/(2n\theta)$.  Simulation studies conducted by \citet{Shilane-nb-2010} demonstrate that the Chi Square approximation will provide reasonably good coverage when the ratio statistic is reasonably close to 1, such as values between $2/3$ and 1.5.  Moreover, the ratio statistic provides information about whether the Chi Square interval is too narrow or too wide.  When the ratio statistic exceeds 1, the interval is too wide; and when the ratio is less than 1, it is too narrow.  This information may be used to compare the results of other procedures even when the Chi Square method performs poorly.  However, it should be emphasized that the Chi Square approximation should be limited in its applications, and asymptotically it will severely over--cover the mean.

\subsection{Bernstein's Inequality}\label{berninequality_P4}

\noindent At small sample sizes and high dispersion, parametric methods that construct confidence intervals by inverting hypothesis testing procedures \citep{ClopperPearson, Sterne, CrowGardner, CasellaBerger} may be inadequate. Tail probability bounds provide an alternative methodology that typically rely upon more mild assumptions about the data.  Bounds such as Bernstein's Inequality \citep{Bernstein}, Bennett's Inequality \citep{Bennett1962, Bennett1963}, or methods based on the work of \citet{Hoeffding} and Berry-Esseen \citep{Berry, Esseen1942, Esseen1956, vanBeek} may be employed.  \citet{rosenblum&vdlaan} apply these bounds to produce confidence intervals based on the estimators' empirical influence curves.\\

\noindent In the Negative Binomial setting, \citet{Shilane-nb-2010} adapted Bernstein's Inequality to provide an improvement over a naive confidence interval.  The method requires only independent data with finite variance and imposes a heuristic assumption of boundedness in a range $(a,b)\in\mathbb{R}$.  Under these conditions, a variant \citep{vdLaanRubinTechRep180} of Bernstein's Inequality may be applied to generate the following confidence interval for $\mu$:

\begin{equation}\label{bernci}
\bar X \pm \frac{\frac{-2}{3}(b-a)\log(\alpha/2) +\sqrt{\frac{4}{9}(b-a)^2[\log(\alpha/2)]^2 - 8n\sigma^2\log(\alpha/2)}}{2n}.
\end{equation}

\noindent In addition to estimating the variance $\sigma^2$ with $s^2$, the Bernstein confidence interval requires the selection of a bounding range $(a,b)$.  Since Negative Binomial variables are non--negative, $a=0$ is a natural choice.  However these data are unbounded above, so any finite selection of $b$ will impose a heuristic bound on the data.  As a default, one may select the sample's maximum value or some multiple thereof.  \citet{Shilane-nb-2010} also considered a variant of Bernstein's Inequality for unbounded data.  However, they found the method to be impractical due to the extremely conservative nature of the tail bound in this setting.\\

\noindent The bounded variant of Bernstein's Inequality improved upon the alternative methods in coverage for small sample sizes and high dispersions.  However, its confidence intervals were far wider and more variable than the other candidates' results.  While this is an improvement over a naive interval for the data's mean (e.g. all values between zero and the sample's maximum), the Bernstein method lacks the interpretative quality provided by parametric approximations.  Furthermore, Bernstein's Inequality is a conservative bound, so asymptotically the method will significantly over--cover the mean.

\section{Growth Estimators for $\mu$}\label{growth}

\noindent The Gamma, Chi Square, and Normal approximations all seek to utilize the existing data to generate an inference for $\mu$ under parametric assumptions about the distribution of $\bar X$.  However, none of these techniques directly considers the data's relative sparsity under high dispersion.  When the probability of a zero value is high, many samples of data will include more than this expected proportion due to chance error.  This over--representation of zeros exacerbates the difficulty of estimation in what is already a sparse data setting.  For instance, in the case of $\mu=10$ and $\theta=0.025$, we expect $85.3\%$ of the sample to be zeros.  If the underlying experiment were repeated a large number of times, then roughly half of the samples would have more than $85.3\%$ zeros.  Furthermore, if $n=30$ samples are drawn, a simulation experiment suggests that the resulting sample mean will be less than $\mu$ roughly $66\%$ of the time.\\

\noindent We propose \emph{growth estimators} for $\mu$ as a method of accounting for the potential over--representation of zeros in the data set.  We will construct growth estimators using two separate procedures:  adjusting the mean through a multiplicative growth factor and direct removal of some zero values from the data.  The details of these procedures are provided in Sections \ref{GBR} and \ref{GBA}.\\

\noindent These growth procedures are motivated by shrinkage estimators.  For instance, in constructing a confidence interval for the Binomial propotion $p$, \citet{agresti} suggest augmenting the existing data with two additional successes and two additional failures.  A Normal approximation confidence interval for $p$ based on these augmented data can be shown to perform measurably better than the Normal approximation alone.  The method of \citet{agresti} effectively shrinks the estimate of $p$ toward the value 0.5 by adding a small amount of data.  By contrast, we seek to grow the estimate of the Negative Binomial mean $\mu$ by removing a small amount of zero--valued data.

\subsection{Growth By Adjustment (GBA)}\label{GBA}

Suppose we believe the data set contains approximately $k\in\mathbb{R}^+$ too many zero values.  The removal of these zeros is tantamount to re--weighting the sample mean by the \emph{growth factor}

\begin{equation}\label{g}
G = \frac{n}{n-k}.
\end{equation}

\noindent Therefore, we estimate $\mu$ with the value

\begin{equation}\label{growtheq}
\tilde X = \frac{1}{n-k}\sum_{i=1}^{n} X_i = \frac{n}{n-k}\bar X = G\bar X.
\end{equation}

\noindent This growth estimator $\tilde X$ has the expected value

\begin{equation}\label{expected}
E[\tilde X] = \frac{n}{n-k} \mu = G\mu,
\end{equation}

\noindent and the standard error is

\begin{equation}\label{sterrGBA}
SD_{\texttt{GBA}}(\tilde X) = \frac{\sqrt{n}}{n-k} \sigma = \left[\frac{n}{n-k}\right]\frac{\sigma}{\sqrt{n}} = G \cdot SD(\bar X).
\end{equation}

\subsection{Growth By Removal (GBR)}\label{GBR}

The GBA method artificially inflates the sample mean by the growth factor $G$.  However, this growth may also be achieved through direct removal of $k$ zeros.  We will refer to this procedure as \emph{Growth By Removal} (GBR).  As in the GBA method, this procedure depends upon the value of $k$.  In general, $k$ should be selected according to a pre--determined rule, and it may not exceed the overall number of zeros in the data set.  In practice, $k$ should be less than this maximum because it is intended to remove only extraneous values.\\

\noindent In the original sample, the data's sum had the expected value $n\mu$.  Since removing zero values does not change this sum, the mean of the $n-k$ remaining values has the expected value $G\mu$, as in Equation (\ref{expected}).  However, the associated standard error is computed differently than in the GBA method.  The standard deviation of the remaining data is centered around $G\bar X$ rather than $\bar X$.  The standard error then divides this quantity by $\sqrt{n-k}$.  Assuming the data $X_1,\dots,X_n$ are sorted in decreasing order, this is given by

\begin{equation}\label{sterrGBR}
SD_{\texttt{GBR}}(\tilde X) =\sqrt{\frac{\displaystyle\sum_{i=1}^{n-k}\left(X_i-\frac{n}{n-k}\bar X\right)^2}{(n-k-1)(n-k)}}.
\end{equation}

\subsection{Convergence to the Normal Approximation}\label{normalconvergence}

\noindent Whether we employ the GBR or GBA methods, the mean may be adjusted by the growth factor $G$.  For any fixed value of $k$, the growth estimator $G\bar X$ will asymptotically converge in sample size to $\bar X$.  Since the Central Limit Theorem applies, we propose a Normal approximation based upon this adjustment.  We will rely upon the unbiased estimate $s^2$ of the variance $\sigma^2$ from the full data set (with no zeros removed). With $z$ defined as the $(1-\alpha/2)$th quantile of the standard Normal distribution, the GBA confidence interval will have the form

\begin{equation}\label{growthciGBA}
\textbf{GBA: } G\bar X \pm zG \frac{s}{\sqrt{n}} = \frac{n}{n-k}\bar X \pm z \frac{\sqrt{n}}{n-k} s.
\end{equation}

\noindent Meanwhile, the GBR method's interval applies its alternative computation of the standard error.  That is,

\begin{equation}\label{growthciGBA}
\textbf{GBR: } G\bar X \pm z \cdot SD_{\texttt{GBR}}(\tilde X) = \frac{n}{n-k}\bar X \pm z \sqrt{\frac{\displaystyle\sum_{i=1}^{n-k}\left(X_i-\frac{n}{n-k}\bar X\right)^2}{(n-k-1)(n-k)}}.
\end{equation}

\subsection{Selection of $k$}\label{kselection}

\noindent Interestingly, the growth estimator adds both bias and variance to the Normal approximation of $\mu$.  The degree of additional error may be controlled through the selection of $k$, the number of zeros to remove from the data set.  We emphasize that this selection should be made with extreme caution.  In Section \ref{sim}, we will explore how the Growth method's coverage probability is impacted by the choice of $k$.  Based upon the results of these simulation experiments, we recommend the following default choices of $k$:

\begin{equation}\label{k}
%$$
k = \begin{cases} \min\left(15, \frac{n}{10}\right), & \text{if } \hat\theta\leq 0.5 \\ 
     \min\left(5, \frac{n}{10}\right), & \text{if } \hat\theta > 0.5 \end{cases}.
%$$
\end{equation}

\noindent The intuition behind these choices is as follows:  at small sample sizes, no more than one zero may be removed per ten data points.  When the dispersion is high ($\theta \leq 0.5$), we allow for more aggressive removal of zeros -- up to a maximum of 15 -- to account for a higher degree of over--representation.  At more moderate dispersions, we limit this removal to no more than 5 zeros.  The value of $\theta$ may be estimated using maximum likelihood estimation or the method of moments when the former is not available.

\section{Simulation Studies}\label{sim}

\subsection{Experimental Design}\label{design}

We designed a simulation experiment to compare the Growth methods to the Bernstein, Gamma, Chi Square, and Normal confidence intervals.  We did not include the bootstrap BCA method in the experiment due to its computational requirements and similarity to the Normal approximation in the simulations of \citet{Shilane-nb-2010}.  We selected an extensive set of parameters and sample sizes, with values displayed in Table \ref{simparams}.  The most extreme case of $\mu=10$ and $\theta=0.025$ would roughly correspond to flipping a coin with a one--fourth of one percent chance of landing heads.  Meanwhile, the most moderate case of $\mu=2$ and $\theta=1$ is equivalent to flipping a coin with a $1/3$ chance of heads.\\

\noindent   Each combination of $\mu$, $\theta$, and $n$ led to a unique and independent simulation experiment.  Each experiment generated 10000 independent size $n$ sets of i.i.d. $NB(\mu,\theta)$ random variables in the \textbf{R} statistical programming language.  On each size $n$ data set, $95\%$ confidence were generated according to each proposed method.  We then approximated the coverage probability of each method by the empirical proportion of confidence intervals containing the true value of $\mu$.\\

\begin{table}[ht]
\begin{center}
\begin{tabular}{|c|l|}
\hline
\textbf{Parameter} & \textbf{Values}\\
\hline
$\mu$ &$\{2, 5,10\}$ \\
\hline
$\theta$ & $\{0.025, 0.05, 0.075,$\\
& $ 0.1, 0.2, 0.3, \dots, 1\}$\\
\hline
$n$ & $\{5,10,15, 20,\dots,250,$\\
& $300, 400, \dots, 1000\}$\\
\hline
$\alpha$ & $0.05$\\
\hline
Trials & $10000$\\
\hline
\end{tabular}
\caption{Parameter values for the simulation
experiments of Section \ref{sim}.}
\label{simparams}
\end{center}
\end{table}

\subsection{Coverage Probability Results}\label{results}

\begin{figure}\begin{center}
  \includegraphics[scale=0.78]{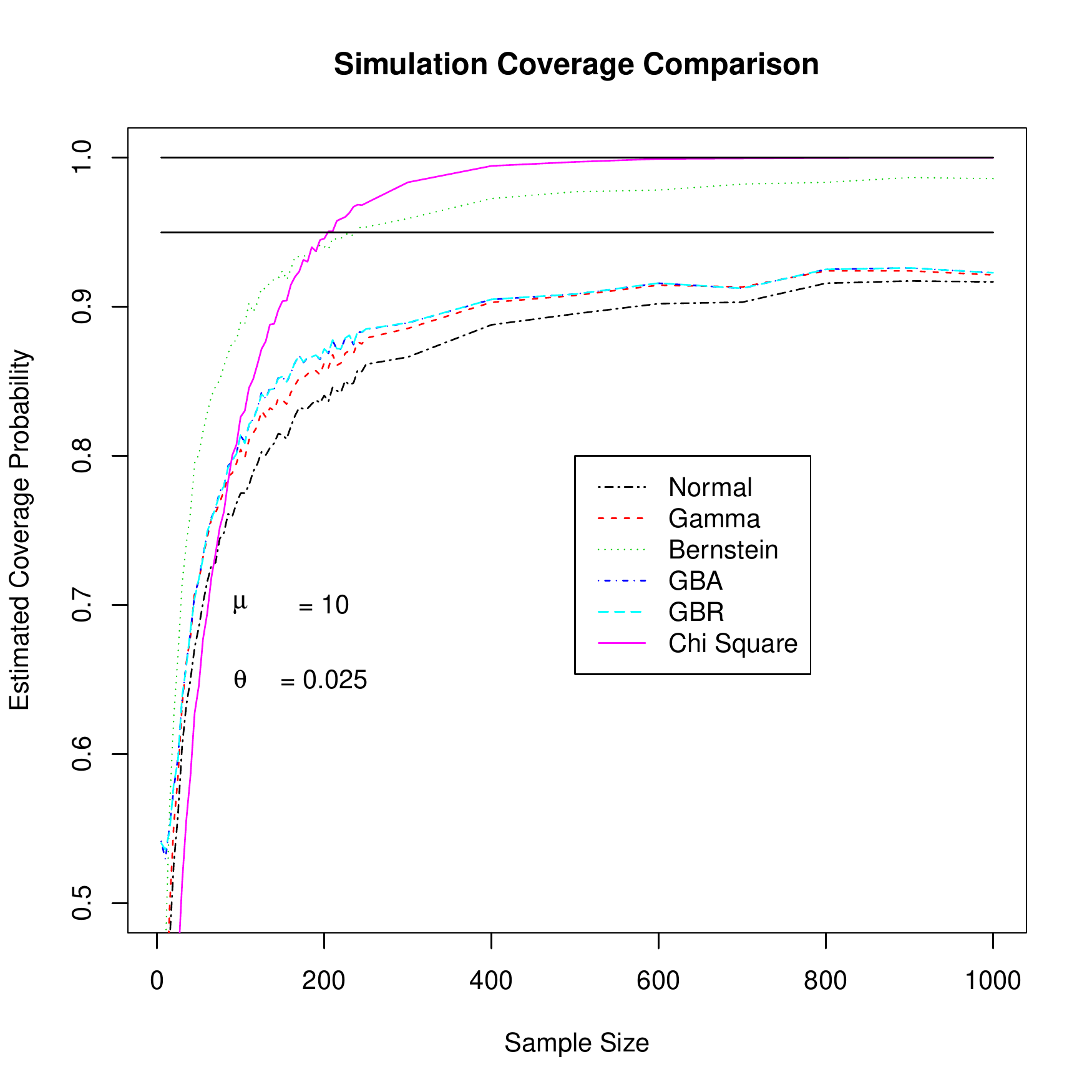}
  \caption{Simulation coverage probabilities for the proposed methods.  With $\mu=10$ and $\theta=0.025$, this represents the most extreme dispersion among all simulation examples.}  \label{mu10theta0025}
  \end{center}
\end{figure}

\begin{figure}\begin{center}
  \includegraphics[scale=0.78]{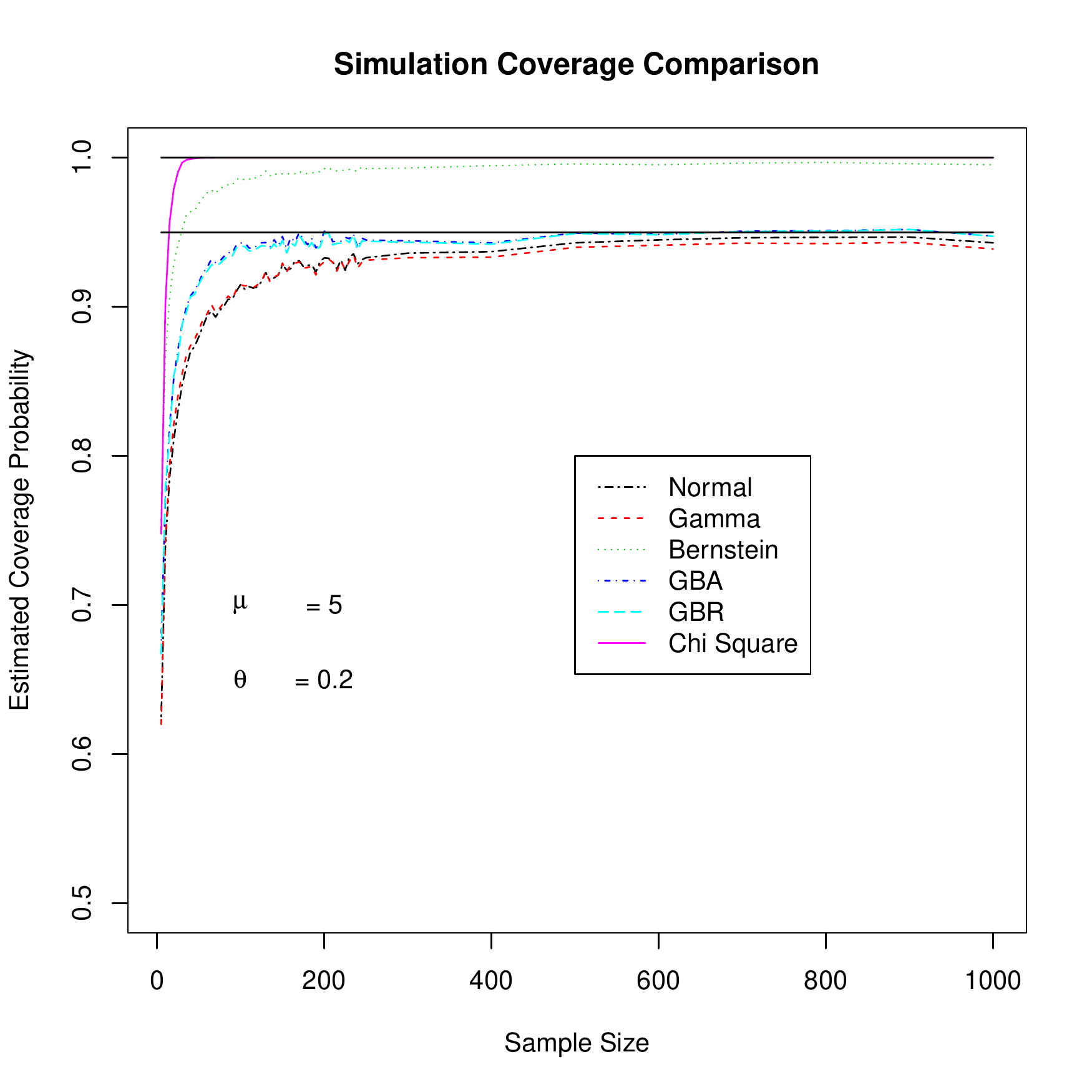}
  \caption{Simulation coverage probabilities for the proposed methods.  With $\mu=5$ and $\theta=0.2$, this represents an intermediate dispersion among the range of simulation examples.}  \label{mu5theta02}
  \end{center}
\end{figure}

\begin{figure}\begin{center}
  \includegraphics[scale=0.78]{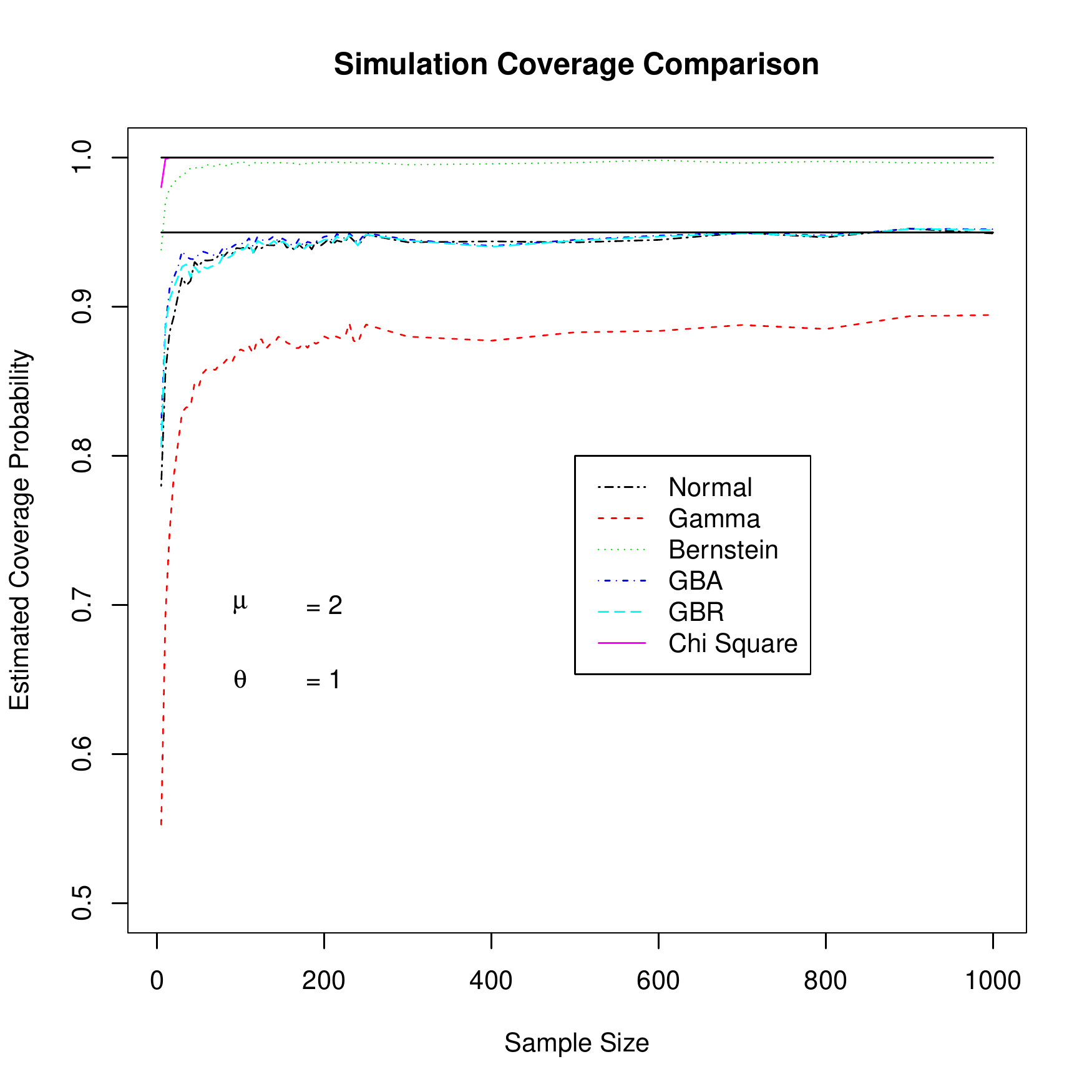}
  \caption{Simulation coverage probabilities for the proposed methods.  With $\mu=2$ and $\theta=1$, this represents the most moderate dispersion among all simulation examples.}  \label{mu2theta1}
  \end{center}
\end{figure}

\noindent Figures \ref{mu10theta0025}, \ref{mu5theta02}, and \ref{mu2theta1} provide examples of the simulation results at high, medium, and low dispersions.  
Each figure displays estimated coverage probabilities for the Normal, Gamma, Chi Square, Bernstein, GBA, and GBR methods as a function of sample size at particular combinations of $\mu$ and $\theta$.  Figure \ref{mu10theta0025} depicts the case of $\mu=10$ and $\theta=0.025$, where the dispersion is the most extreme.  Here even the Bernstein Method requires a significantly large sample size to reach the desired $95\%$ coverage probability.  The Chi Square approximation performs as expected by reaching $95\%$ coverage almost exactly when $\mu=2n\theta$ and over--covering $\mu$ for larger sample sizes.  Because $\theta$ is very small, we expect the Gamma approximation to outperform the Normal.  However, because of the extreme dispersion, both methods require an extremely large sample size to appropriately cover the mean.  Even at $n=250$, the Gamma method only covers $\mu$ $87.91\%$ of the time, and the Normal only reaches a rate of $85.92\%$.  Meanwhile, the Growth methods provide small but steady improvements over the Gamma approximation.\\

\noindent Figure \ref{mu5theta02} displays coverage probabilities for $\mu=5$ and $\theta=0.2$, where the dispersion is moderate and more typical of a Negative Binomial study.  Both the Bernstein and Chi Square methods quickly over--cover the mean, surpassing $95\%$ by $n= 30$ and $n=15$, respectively.  The Gamma and Normal approximations are roughly in equipoise at this moderate level of dispersion.  The Growth method improves upon their coverage by about $3\%$ at sample sizes up to 100 and maintains at least a $1\%$ improvement even out to $n=250$.  Despite the fairly moderate dispersion, the Normal and Gamma approximations appear to require 250 or more data points to approach convergence, whereas the Growth method reaches this point at about $n=100$.\\

\noindent Figure \ref{mu2theta1} displays coverage probabilities in the case of $\mu=2$ and $\theta=1$.  Because the dispersion is quite small, the Negative Binomial model is reasonably close to the Poisson distribution.  We therefore expect the Normal approximation to perform quite well.  Even in this case, the GBA method provides small improvements over the Normal at small and moderate sample sizes.  The GBR method also provides early improvements.  However, its coverage briefly dips below that of the Normal approximation at about $n=50$.   With relatively few zeros removed from the data, the Growth methods and Normal approximation quickly fall into agreement.  By contrast, the Gamma approximation does not perform well in this scenario because its underlying limit theorem requires high dispersion relative to the sample size.  Similarly, the Chi Square and Bernstein methods almost immediately over--cover the mean.

\subsection{The Effect of Misspecified Growth}\label{misspecified}

\noindent The Growth Method simulation results of Section \ref{results} were obtained under the default selections of $k$ given by Equation (\ref{k}).  These recommended settings were obtained through a process of trial and error as applied to the simulation studies.  The intuition of these recommendations is that zero values may be removed more aggressively under higher dispersions.  We believe the extensive range of parameters tested in the simulation study provide reasonable evidence that these recommendations will generalize well.  Other approaches to selecting $k$ may also consider the observed sample mean or alter the gradations by sample size.\\

\begin{figure}\begin{center}
 \includegraphics[scale=0.78]{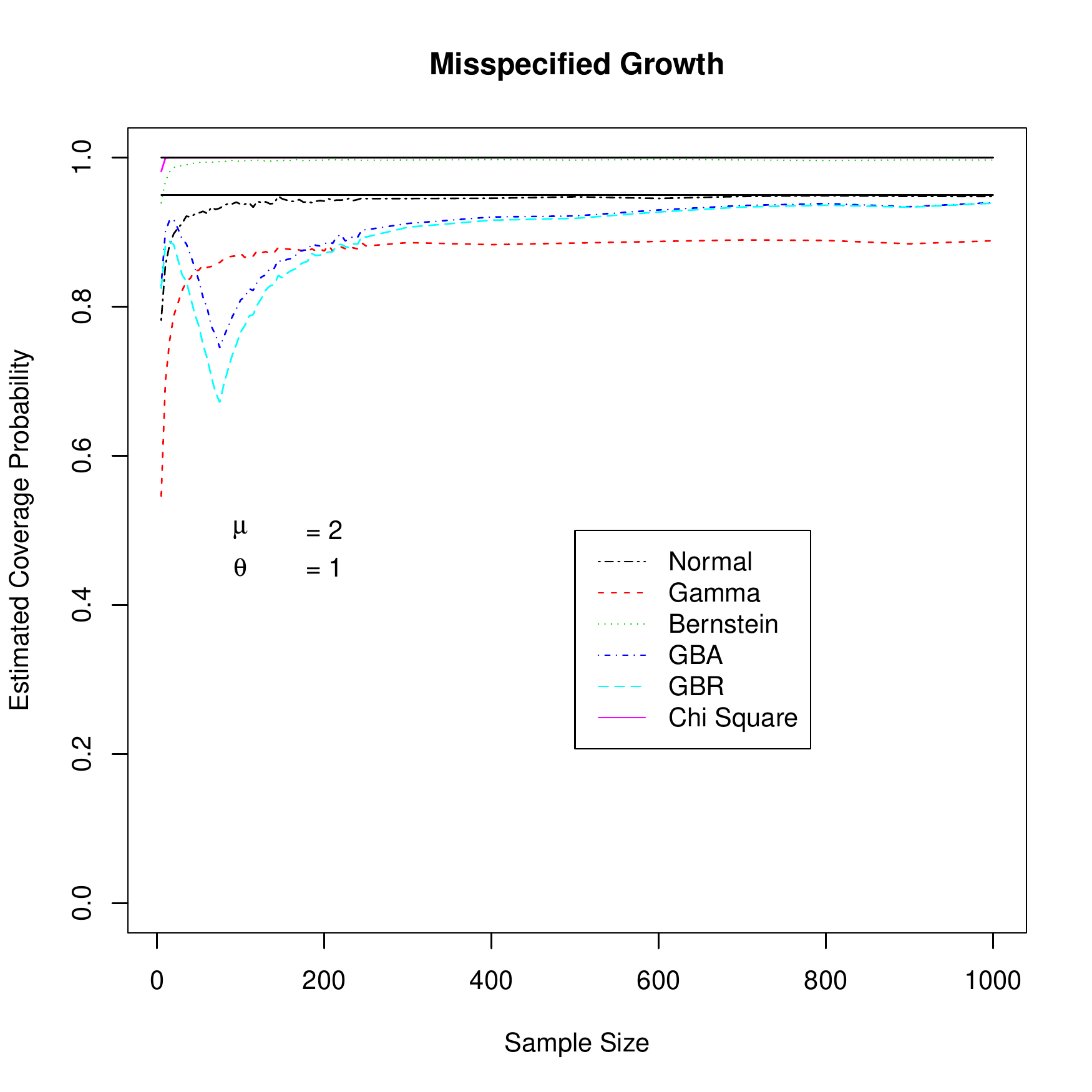}
  \caption{An example of misspecified growth.  Despite low dispersion, zeros were aggressively removed at a rate of one for every 5 data points up to a maximum of 15.  The GBA and GBR methods' coverage probabilities dip significantly until this maximum value is reached and then rebound toward the Normal approximation.}  \label{altk}
  \end{center}
\end{figure}

\noindent We urge the practitioner to exercise caution in selecting how many zeros to remove.  An overzealous selection of $k$ may lead to poor performance in the Growth Method's coverage probability.  As an example, we repeated the simulation study in the case of $\mu=2$ and $\theta=1$ with $k=\min(15, n/5)$ instead of the recommended valued of $k=\min(5, n/10)$.  This aggressive approach removes zeros at double the rate and allows for a maximum that triples the recommendations for the low dispersion case of $\theta=1$.  Figure \ref{altk} displays the consequences of misspecified growth.  Rather than the small improvement over or close agreement to the Normal approximation as seen in Figure \ref{mu2theta1}, the Growth Method decreases significantly in coverage.  This dip in performance continues until the maximum removal is reached at $n=75$.  For larger sample sizes, the Growth Method begins to rebound toward the Normal approximation.  However, this case suggests that the practitioner should be careful not to select a value of $k$ that is too large, especially when the dispersion is mild.\\

\begin{figure}\begin{center}
 \includegraphics[scale=0.78]{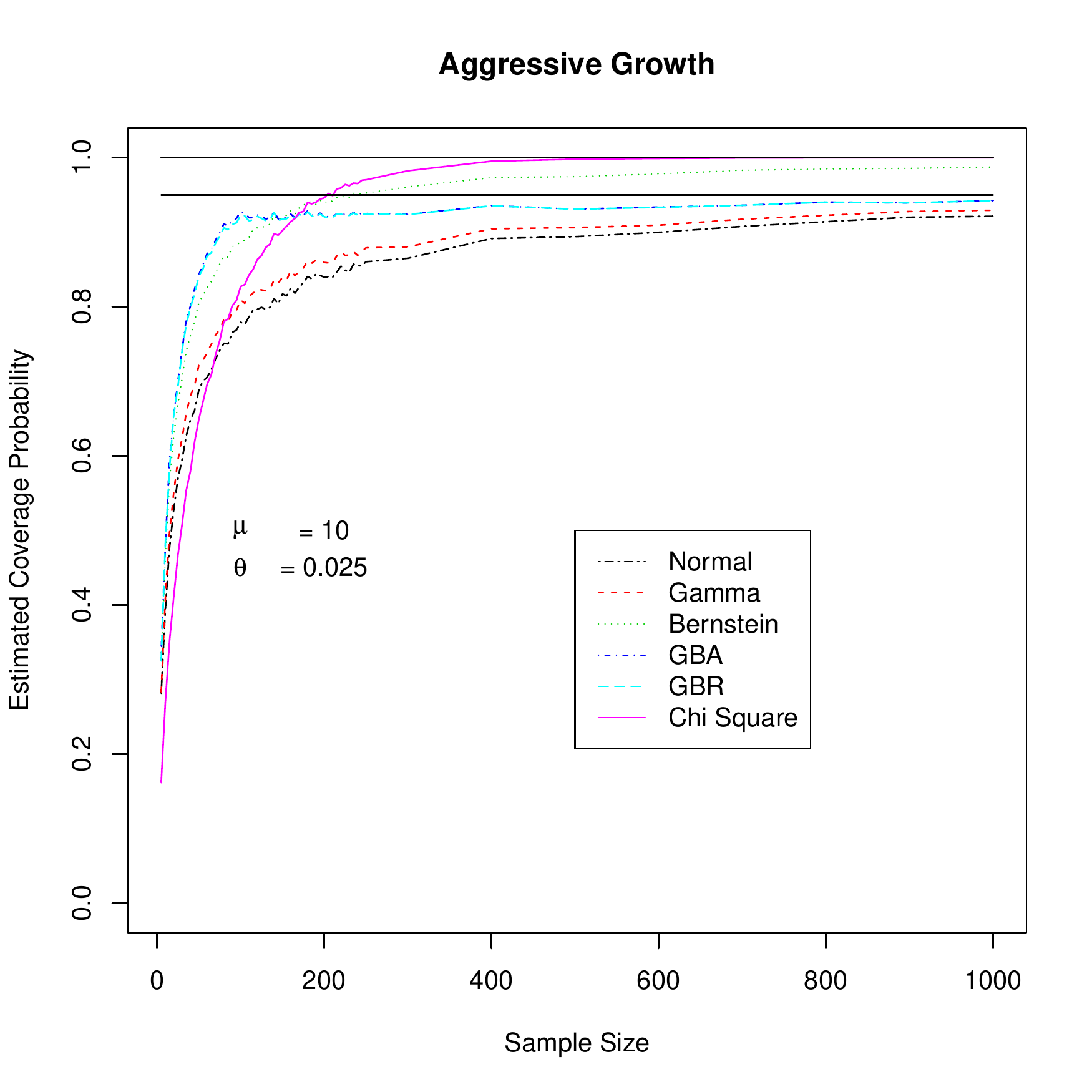}
  \caption{An example of aggressive growth.  Under high dispersion, zeros were aggressively removed at a rate of one for every 2 data points up to a maximum of 50.  The GBA and GBR methods' coverage probabilities accelerate quickly in this setting.}  \label{altk2}
  \end{center}
\end{figure}

\noindent By contrast, higher dispersion settings allow for far more aggressive growth.  We also repeated the simulation study in the case of $\mu=10$ and $\theta=0.025$, the most extreme dispersion considered.  Figure \ref{altk2} depicts the Growth Method's coverage when $k=\min(50, n/2)$, which effectively removes half the data points until sample size 100.  In this example, the Growth method crosses the $90\%$ coverage threshold by $n=80$, a mark not reached by the Gamma or Normal approximations by $n=250$. Indeed, the Growth Method accelerates in coverage even faster than the Bernstein Method.\\

\noindent Overall, the GBA method appears to perform slightly better than the GBR method across all simulation experiments.  This difference may be attributed to the selection of $k$.  The GBR method requires an integer value so that exactly $k$ zeros may be removed.  By contrast, the GBA method allows for adjustments using continuous values of $k$.  The recommended selection procedure of Equation (\ref{k}) allows for fractional proportions of the overall sample size.  This additional fraction allows for increased growth, which in turn leads to improved coverage.  The GBA and GBR methods also differ in the computation of their standard errors.  It appears that these standard errors are largely similar in value.  We will substantiate this claim further in Section \ref{cilength}.

\subsection{Confidence Interval Length Considerations}\label{cilength}

\begin{figure}\begin{center}
  \includegraphics[scale=0.78]{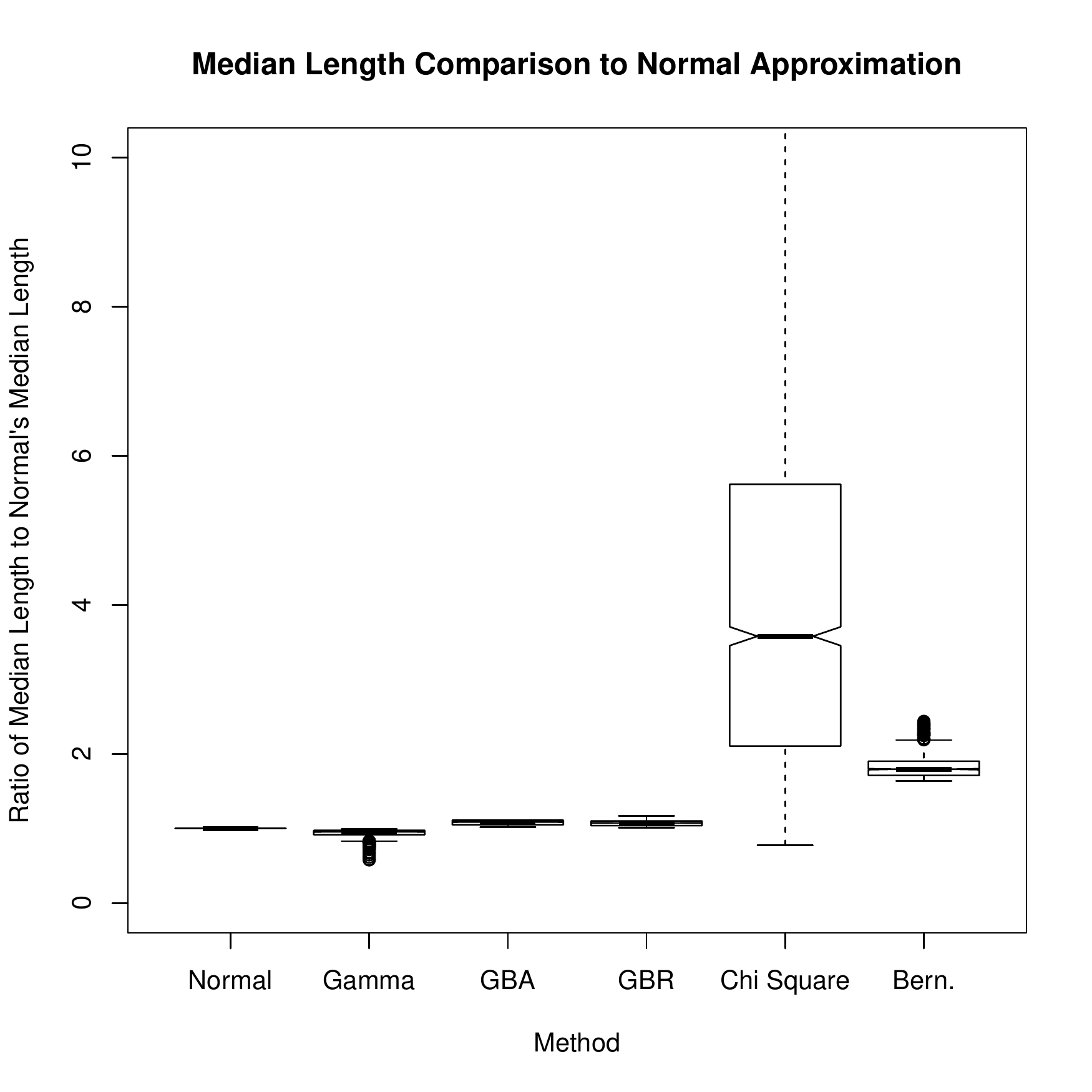}
  \caption{A comparison of median confidence interval length standardized by the Normal approximation's median values.  Each simulation experiment computed the median interval length of each method.  The values depicted are the ratios of these medians to that of the Normal approximation.}  \label{medratios}
  \end{center}
\end{figure}

\begin{figure}\begin{center}
  \includegraphics[scale=0.78]{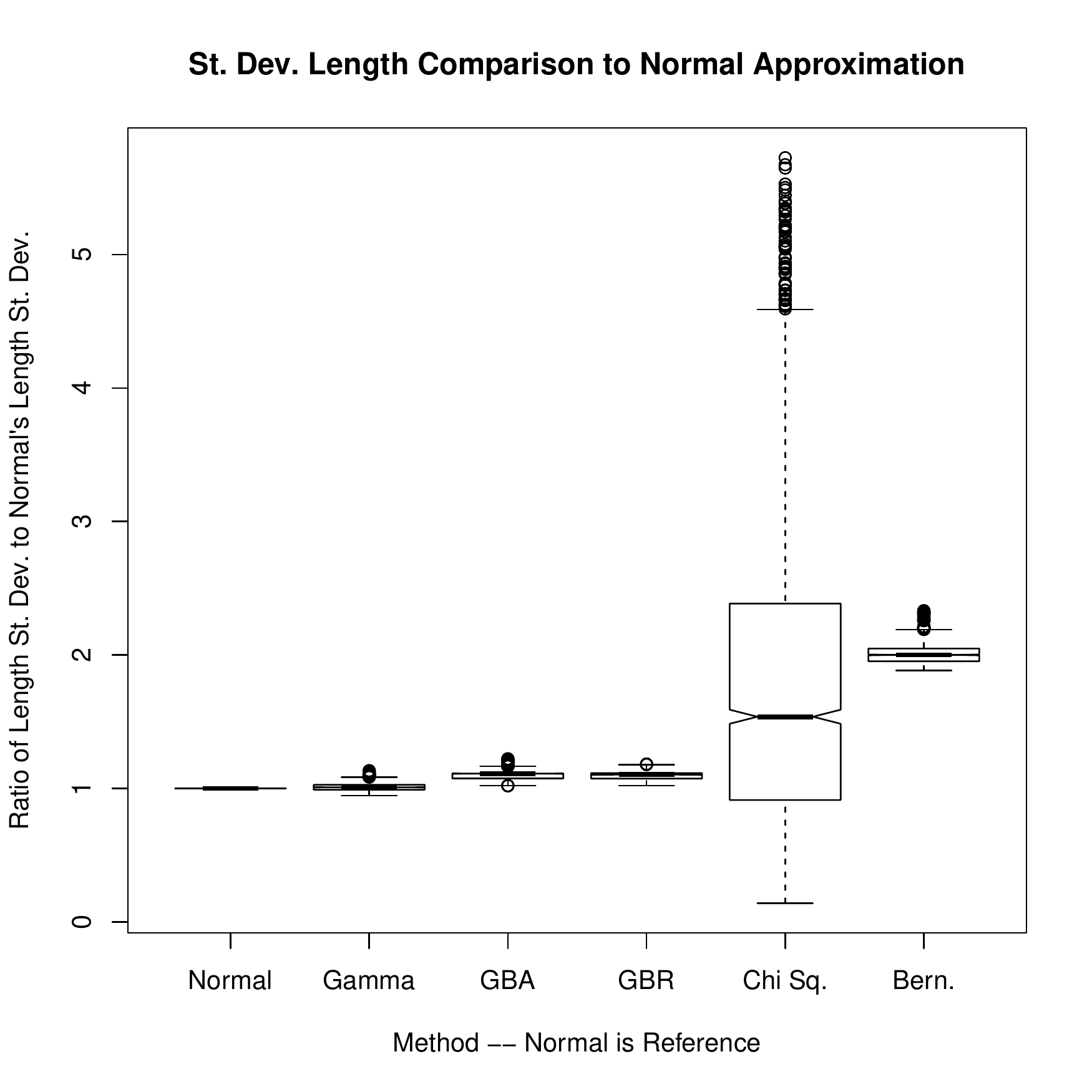}
  \caption{A comparison of confidence interval length stability.  Each simulation experiment computed the standard deviation of interval length of each method.  The values depicted are the ratios of these standard deviations to that of the Normal approximation.}  \label{sdratios}
  \end{center}
\end{figure}

\noindent We have adopted coverage probability as our preferred metric of a confidence interval's quality.  However, the length of these intervals is an important secondary consideration.  Shorter intervals suggest greater precision in the estimator when the coverage probability is approximately equal.  Figures \ref{medratios} and \ref{sdratios} provide a comparison of the proposed methods' lengths.  In each simulation experiment, we recorded the median and standard deviation of length for the 10000 confidence intervals generated by each method.   We then computed the ratio of each method's median length to that of the Normal approximation in each experiment.  Figure \ref{medratios} displays the distribution of these ratios, and Figure \ref{sdratios} depicts the ratio of the standard deviation of length.\\  

\noindent In general, it appears that the Bernstein method produces confidence intervals that are typically a factor of 1.8 longer than the corresponding Normal approximation.  Because the GBA and GBR methods usually remove about one zero per ten data points, their median lengths were typically a factor of $\frac{10}{9}$ larger than the Normal approximation's interval.  Likewise, the same growth factor applies to the standard deviations of length in Figure \ref{sdratios}.  The Bernstein Method has a length variability that is roughly double that of the Normal approximation.  This increased variability of length leads to the over--coverage in the method, as extremely long confidence intervals are far more likely to contain the mean.  By contrast, the Gamma method typically produces a confidence interval that is $95\%$ the length of the Normal approximation.  This length depends on the degree of dispersion, with higher lengths (and improved coverage) occurring at smaller values of $\theta$.

\section{Discussion}\label{conclusion}

\noindent The proposed growth methods provide improved confidence intervals for the mean of Negative Binomial random variables with unknown dispersion.  Removing a small number of zeros from the data set bolsters the coverage probability at small and moderate sample sizes.  Asymptotically, the GBA and GBR methods converge to the Normal approximation.  Overall, applying a growth estimator produces intervals that are longer and more variable than the Normal approximation.  The degree of increase may be controlled through a selection of the growth factor $G$, or, equivalently, the removal factor $k$.  This selection depends on the sample size and degree of dispersion in the data.  We emphasize that the number of zeros $k$ to remove should be selected cautiously to prevent coverage drop--offs such as the example depicted in Figure \ref{altk}.\\

\noindent The previous work of \citet{Shilane-nb-2010} provided a piecewise solution to performing inference on the Negative Binomial mean.  The Gamma, Chi Square, and Normal approximations performed well in largely complimentary settings, and Bernstein's Inequality was used at smaller sample sizes and high dispersions.  However, we have demonstrated that the GBA and GBR methods can perform well in a wide variety of settings.  Using the relatively simple guidelines for selecting $k$ in Equation \ref{k}, these procedures outperformed the parametric approximations at both high and low dispersions.  Because it allows for continuous values of $k$, the GBA method generally provided small improvements over the GBR results.  A tail probability bound such as Bernstein's Inequality may still be considered at very small sample sizes and extremely high dispersions, but the Growth Method accelerates quickly as a function of sample size.\\

\noindent Future work on this problem may provide more solid theory for how the value of $k$ should be selected.  Growth estimators add both bias and variance to the Normal approximation, so a traditional bias--variance trade-off calculation does not apply.  Indeed, a criterion such as the mean squared error would be optimized with the selection of $k=0$, which is equivalent to the Normal approximation.  The cautionary tale depicted in Figure \ref{altk} suggests that coverage is optimized at some intermediate value of $k$.  However, the analytic coverage probability calculation is intractable because it depends upon the permutation distribution of the Negative Binomial sample.\\

\noindent The Growth Method may be extended to other sparse estimation problems.  \citet{Shilane-nb2-2011} previously examined the quality of the Normal approximation in two--sample Negative Binomial inference, and growth estimators may be easily adapted to this setting.  We are also presently examining an application of the growth method for estimating the mean of Gamma random variables.

\bibliographystyle{chicago}
\bibliography{nb}

\end{document}